\documentclass[
]{ceurart}

\usepackage{booktabs}
\usepackage{anyfontsize}
\usepackage{amsfonts}
\usepackage{amsmath}
\usepackage{graphicx}
\usepackage{balance}
\usepackage[T1]{fontenc}
\usepackage{subfigure}
\usepackage{algorithm2e}
\usepackage{amsthm}
\usepackage{hyperref}
\usepackage{xurl}
\usepackage{pifont}
\usepackage{multirow}
\usepackage{array}
\usepackage[inline]{enumitem}
\usepackage{bbding}
\usepackage{wasysym}

\usepackage[skip=0.5\baselineskip]{caption}

\usepackage{listings}

\lstdefinelanguage{SPARQL}{
  morekeywords={
    SELECT, CONSTRUCT, ASK, DESCRIBE,
    WHERE, FILTER, OPTIONAL, GRAPH,
    UNION, PREFIX, BASE, LIMIT, OFFSET, ORDER, BY, ASC, DESC,
    DISTINCT, REDUCED, FROM, NAMED
  },
  sensitive=true,
  morecomment=[l]{\#},
  morestring=[b]",
}

\usepackage{listings}
\begin{document}

\copyrightyear{2026}
\copyrightclause{Copyright for this paper by its authors. Use permitted under Creative Commons License Attribution 4.0 International (CC BY 4.0).}

\conference{Second International TEXT2SPARQL Challenge, Co-Located with Text2KG at ESWC26, May 10, 2026, Dubrovnik, Croatia.}

\title{SPARQL-LLM in the TEXT2SPARQL 2026 Challenge}

\author[1]{Panayiotis Smeros}[
email=panayiotis.smeros@sib.swiss]
\cormark[1]
\fnmark[1]

\author[1]{Vincent Emonet}[
email=vincent.emonet@sib.swiss]
\fnmark[1]

\author[1,2]{Ruijie Wang}[
email=ruijie.wang@sib.swiss]

\author[1]{Ana-Claudia Sima}[
email=ana-claudia.sima@sib.swiss]

\author[1]{Tarcisio {Mendes de Farias}}[
email=tarcisio.mendes@sib.swiss]

\address[1]{SIB Swiss Institute of Bioinformatics}
\address[2]{University of Zurich}

\cortext[1]{Corresponding author.}
\fntext[1]{These authors contributed equally.}

\begin{abstract}
Querying distributed data represented as Knowledge Graphs (KGs) via SPARQL remains a difficult task, requiring fluency in query syntax, endpoint specifications, and schema details. Knowledge Graph Question Answering (KGQA) systems address this by automatically translating natural language into SPARQL, and recent advances in Large Language Models (LLMs) have made this translation increasingly feasible. In this paper, we describe SPARQL-LLM, our open-source KGQA system, in the context of the TEXT2SPARQL 2026 challenge. In this challenge, SPARQL-LLM ranks first overall among eight competing approaches across all datasets, achieving up to 59\% higher accuracy than the second-best approach and supporting multilingual question answering in English, German, and Spanish. SPARQL-LLM is publicly released at \url{https://github.com/sib-swiss/sparql-llm} and deployed in production over decentralized bioinformatics KGs at \url{https://www.expasy.org/chat}.
\end{abstract}


\maketitle

\section{Introduction}

Accessing distributed data in modern web ecosystems is a persistent challenge, compounded by heterogeneous formats, schemas, and access methods. Representing such data as RDF Knowledge Graphs (KGs) has emerged as a possible practical solution; however, querying this data via SPARQL remains a difficult task \cite{DBLP:journals/bmcbi/McCarthyVW12}, requiring fluency in query syntax, endpoint specifications, schema details, and cross-dataset relationships. KGQA systems aim to address this by automatically translating natural language into SPARQL \cite{DBLP:conf/ijcai/LanHJ0ZW21,diefenbach2018core}, lowering the barrier to Knowledge Graph access and accelerating data discovery. Recent advances in Large Language Models (LLMs) have further strengthened this direction, enabling direct text-to-SPARQL translation \cite{dabramo-etal-2025-investigating}. 

In this context, the TEXT2SPARQL Challenge\footnote{\url{https://text2sparql.aksw.org}}, which was first launched in 2025, introduced a multi-dataset and multi-lingual text-to-SPARQL benchmark with the overall goal of challenging the participating systems to improve the accessibility of Knowledge Graphs through more intuitive querying interfaces.
After a successful first edition with six participating approaches, TEXT2SPARQL was relaunched in 2026 with a second edition. 
In the 2026 edition, eight approaches competed, among them \emph{SPARQL-LLM}~\cite{smeros2025sparql}, the system we present in this paper.

The key takeaways of our participation in the challenge are summarized as follows:

\begin{itemize}
\item \emph{SPARQL-LLM} ranks first overall among eight competing approaches across all categories, achieving up to $59\%$ higher accuracy than the second best approach;
\item \emph{SPARQL-LLM} supports multilingual question answering in English, German, and Spanish;
\item \emph{SPARQL-LLM} is released as an open-source project at \url{https://github.com/sib-swiss/sparql-llm} and deployed over decentralized bioinformatics KGs at \url{https://www.expasy.org/chat}.
\end{itemize}

\section{Related Work}

In the 2025 edition of the TEXT2SPARQL challenge \cite{text2sparql2025}, several systems leveraging LLMs were submitted, which can be broadly organized by their main strategy: agent-based exploration, in which the LLM iteratively invokes tools to discover the schema of the KG \cite{ref:Dorsch2025,ref:Brei2025,ref:Perevalov2025}; fine-tuning of autoregressive models for query generation \cite{ref:Soru2025}; and RAG-based context construction, injecting retrieved example question-query pairs, data shapes, and ontology fragments into the prompt \cite{ref:Wardenga2025,ref:Berezin2025}.
Similarly, in the 2026 edition, some of these systems evolved, while new systems, including SPARQL-LLM \cite{smeros2025sparql}, participated for the first time.
Notably, the Liber‑AI team, as part of their project named Neural SPARQL Machines, proposed two new approaches based on the Qwen and Claude models. 
Additionally, Iris proposed another system using the Qwen model, the InfAI team introduced three closely related variants of ARUQULA, and a team from the University of Freiburg proposed GRASP. 


While the agentic methods proposed by these related approaches have some advantages, since accessing complex tools allows them to adapt to arbitrary queries, they also face critical limitations. 
First, efficiency remains a challenge; each new question often requires rediscovering schema patterns from scratch, resulting in long runtimes (e.g., dozens of tool calls) and high costs due to token usage. 
Second, complex query patterns
remain difficult to handle; agents may fail to converge or require many iterations. 
SPARQL-LLM, on the other hand, follows a retrieval-based approach, aiming for real-time query generation with minimal interactions with the LLM.
By assembling the context ahead of query time, it avoids both limitations by construction: the number of LLM interactions per question is fixed rather than data-dependent, and there is no agent loop that can fail to converge.
In fact, in our experimental evaluation, we showcase that even with the real-time requirement, we can still achieve remarkable performance with respect to the correctness of the generated queries.

\section{Experimental Evaluation}
In this section, we describe the experimental evaluation framework, SPARQL-LLM and the other competing systems, and the corresponding results of the 2026 edition of the TEXT2SPARQL challenge.

\subsection{Experimental Setup}

Similarly to the 2025 edition of the challenge, the 2026 edition also comprises two subtasks: 
\begin{enumerate*}[label=\roman*)]
\item a task with a well-known and well-documented KG (DBpedia), consisting of $50$ natural language questions in English and Spanish, and 
\item a task with an unknown and undocumented KG regarding a private corporation (named Corporate), consisting of $50$ questions in English and German.
\end{enumerate*}

\subsection{SPARQL-LLM}

\begin{table}[t]
\caption{
LLM configuration of SPARQL-LLM; a detailed description of its architecture is provided in \cite{smeros2025sparql}.}
\label{table:LLM_configuration}
\begin{tabular}{r|ccccccc}
System & \shortstack{LLM\\Provider} & Version & Context & Temperature & \shortstack{Max\\Tokens} & Seed & \shortstack{Knowledge\\Cutoff} \\
\hline
$SPARQL-LLM$ & OpenAI & \texttt{GPT-5.2} & 400K & 0 & 16K & 42 & 08/2025 \\
\end{tabular}
\end{table}

A detailed description of the architecture of SPARQL-LLM is provided in \cite{smeros2025sparql}; specifically for this challenge, our LLM configuration is presented in Table~\ref{table:LLM_configuration}.
Furthermore, we provide two main types of context to SPARQL-LLM:
\begin{enumerate*}[label=\roman*)]
\item existing examples of pairs of natural language questions and SPARQL queries, and
\item schema information of the SPARQL endpoints.
\end{enumerate*}

For DBpedia, we use the example sets proposed by the challenge organizers, which consist of the following:
\begin{enumerate*}[label=\roman*)]
\item LC-QuAD \cite{DBLP:conf/semweb/TrivediMDL17}, a template-generated KGQA corpus,
\item QALD-9+ \cite{DBLP:conf/semco/PerevalovDUB22}, an aggregate of nine editions of the KGQA corpus, QALD, and
\item the test set of the 2025 edition of the challenge, which consists of $100$ natural language questions in English and Spanish.
\end{enumerate*}

As for Corporate examples, since the KG is completely unknown and undocumented, and the provided ontology is relatively small---13 classes and 30 properties, we utilized an LLM, appropriately prompted to generate pairs of natural language questions and SPARQL queries\footnote{The prompt used for generating examples: 
\textit{You are an assistant that helps users formulate pairs of Natural Language Questions that correspond to SPARQL Queries.
Your role is to generate \{number\} example pairs based on the provided ontology consisting of 13 classes and 30 properties: \{ontology\}}
}.
The LLM we utilized (DeepSeek-V3) was released before the publication of the test questions of the challenge; thus, it had no access to these questions during its training.
%
%
Finally, as an additional verification step, we run the generated queries against the SPARQL endpoint of the Corporate KG and discard queries with no results.
Examples of such queries include: ``List products with missing price information'' and ``Show agents with multiple areas of expertise'', which cover the basic SPARQL functionality and the basic knowledge of the KG.
Similarly to DBpedia, we also provided as examples the test set of the 2025 edition of the challenge, which consists of $50$ natural language questions in English.


\subsection{Results}
\begin{figure}[t]
    \centering
    \includegraphics[width=\linewidth]{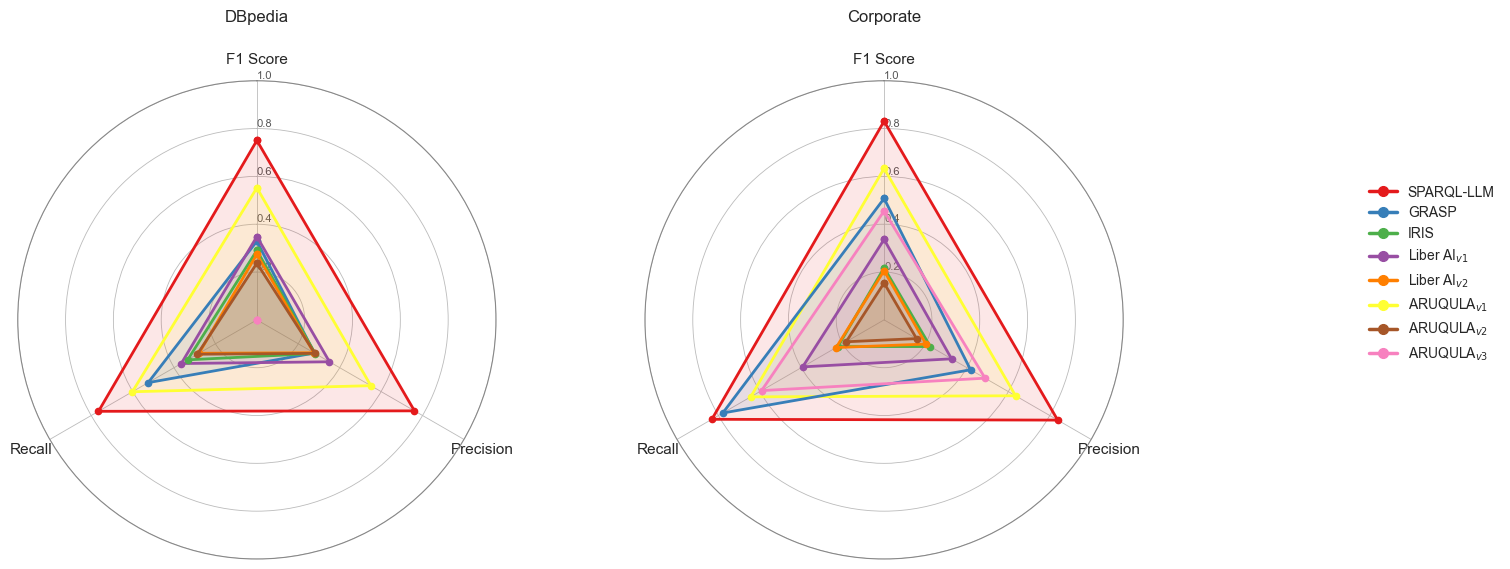}
    \caption{Visualization of the overall results of the experiments produced by \emph{T2S-Metrics} \cite{taghzouti2026t2s}. As we observe, SPARQL-LLM ranks first for both DBpedia and Corporate, with significant margin from the second best system.
    }
    \label{fig:overall_results}
\end{figure}

As shown in Figure~\ref{fig:overall_results}, SPARQL-LLM is evaluated against four other systems, two of which were submitted with multiple variants to the challenge. 
In both DBpedia and Corporate subtasks, SPARQL-LLM showcases a substantial increase in F1 Score relative to the second-best performing system.
In Tables~\ref{table:results_db26_english_2026},~\ref{table:results_db26_spanish_2026},~\ref{table:results_ck26_english_2026},~\ref{table:results_ck26_german_2026}, we present the
breakdown of the results by dataset and by language.
SPARQL-LLM achieves the best result in all subtasks, as illustrated in the result tables, which demonstrates the robustness of the system and of the overall results.
Indicatively, SPARQL-LLM achieves the best overall improvement for DBpedia in Spanish, a $59\%$ higher F1 Score compared to the second-best system.

Finally, the total cost of all the LLM calls of SPARQL-LLM for TEXT2SPARQL 2026 was less than $\$3$, confirming that it is not only the most accurate system but also a system with minimal operational cost.

\begin{table}[ht]
\caption{
English DBpedia results: SPARQL-LLM achieves $17\%$ higher F1 Score than the second-best system (ARUQULA).
}
\label{table:results_db26_english_2026}
\begin{tabular}{r|ccc}
System & F1-score & Recall & Precision \\
\hline
$SPARQL\text{-}LLM$   & \textbf{.7192} & \textbf{.7359} & \textbf{.7292} \\
$GRASP$               & .3296 & .4964 & .2918 \\
$IRIS$                & .2897 & .3312 & .2795 \\
$Liber\ AI_{v1}$      & .3235 & .3360 & .3360 \\
$Liber\ AI_{v2}$      & .3129 & .3199 & .3110 \\
$ARUQULA_{v1}$        & .6142 & .6498 & .6211 \\
$ARUQULA_{v2}$        & .2539 & .3026 & .2767 \\
$ARUQULA_{v3}$        & .0000 & .0000 & .0000 \\
\end{tabular}
\end{table}

\begin{table}[ht]
\caption{
Spanish DBpedia results: SPARQL-LLM achieves $59\%$ higher F1 Score than the second-best system (ARUQULA).
}
\label{table:results_db26_spanish_2026}
\begin{tabular}{r|ccc}
System & F1-score & Recall & Precision \\
\hline
$SPARQL\text{-}LLM$   & \textbf{.7810} & \textbf{.7936} & \textbf{.7927} \\
$GRASP$               & .3288 & .5536 & .2632 \\
$IRIS$                & .2914 & .3374 & .2837 \\
$Liber\ AI_{v1}$      & .3730 & .3943 & .3653 \\
$Liber\ AI_{v2}$      & .2361 & .2390 & .2385 \\
$ARUQULA_{v1}$        & .4904 & .5526 & .4798 \\
$ARUQULA_{v2}$        & .2200 & .2722 & .2835 \\
$ARUQULA_{v3}$        & .0000 & .0000 & .0000 \\
\end{tabular}
\end{table}

\begin{table}[ht]
\caption{
English Corporate results: SPARQL-LLM achieves $33\%$ higher F1 Score than the second-best system (ARUQULA).
}
\label{table:results_ck26_english_2026}
\begin{tabular}{r|ccccc}
System & F1/NDCG & Recall & Precision & F1 & NDCG \\
\hline
$SPARQL\text{-}LLM$   & \textbf{.8539} & \textbf{.8467} & \textbf{.8649} & \textbf{.8510} & \textbf{1.0000} \\
$GRASP$               & .5319 & .7980 & .4373 & .5262 & .7309 \\
$IRIS$                & .2895 & .2923 & .3045 & .2896 & .3482 \\
$Liber\ AI_{v1}$      & .3208 & .3662 & .3151 & .3215 & .3482 \\
$Liber\ AI_{v2}$      & .1745 & .2078 & .1660 & .1724 & .3316 \\
$ARUQULA_{v1}$        & .6409 & .6606 & .6428 & .6437 & .5000 \\
$ARUQULA_{v2}$        & .2067 & .2129 & .2061 & .1937 & .8485 \\
$ARUQULA_{v3}$        & .5211 & .6836 & .5524 & .5279 & .2760 \\
\end{tabular}
\end{table}

\begin{table}[ht]
\caption{
German Corporate results: SPARQL-LLM achieves $30\%$ higher F1 Score than the second-best system (ARUQULA).
}
\label{table:results_ck26_german_2026}
\begin{tabular}{r|ccccc}
System & F1/NDCG & Recall & Precision & F1 & NDCG \\
\hline
$SPARQL\text{-}LLM$   & \textbf{.8141} & \textbf{.8166} & \textbf{.8117} & \textbf{.8104} & \textbf{1.0000} \\
$GRASP$               & .4993 & .7598 & .3979 & .4892 & .8461 \\
$IRIS$                & .1432 & .1541 & .1421 & .1460 & .0000 \\
$Liber\ AI_{v1}$      & .3497 & .4200 & .3368 & .3510 & .3482 \\
$Liber\ AI_{v2}$      & .2399 & .2553 & .2385 & .2390 & .3482 \\
$ARUQULA_{v1}$        & .6242 & .6273 & .6268 & .6230 & .7686 \\
$ARUQULA_{v2}$        & .1125 & .1536 & .1076 & .1148 & .0000 \\
$ARUQULA_{v3}$        & .3792 & .4998 & .4230 & .3827 & .2311 \\
\end{tabular}
\end{table}

\section{Conclusion}
We presented SPARQL-LLM, the top-ranked system in the TEXT2SPARQL 2026 challenge. 
SPARQL-LLM leverages lightweight metadata from SPARQL endpoints to construct an optimized context for the task of translating (multilingual) natural language questions to SPARQL queries. SPARQL-LLM outperformed all the other systems participating in the challenge, achieving the best accuracy at minimal operational cost. The system is available as an open-source project at \url{https://github.com/sib-swiss/sparql-llm}. 

\section*{Acknowledgments}

This work was funded by the SNSF MetaboLinkAI project, grant: 10.002.786 and by SERI via the HORIZON-INFRA-2024-EOSC-01-05 project EOSC Data Commons under grant agreement 101188179. 

\section*{Declaration on Generative AI}
During the preparation of this paper, the authors used Claude (Anthropic)  for grammar, spelling, paraphrasing, and stylistic edits. After using these services, the authors reviewed and edited the content as needed; hence, the authors take full responsibility of the content of this paper.

\bibliography{bibliography}


\end{document}